%% file: paper.tex
\documentclass[twocolumn,showpacs,preprintnumbers,amsmath,amssymb]{revtex4}
\usepackage{dcolumn}
\usepackage{bm}

\usepackage{amsmath}
\usepackage{amsfonts}   
\usepackage{graphics}  
\usepackage{psfrag} 
\usepackage{graphicx} 
\usepackage{epsfig}    
\usepackage{rotating}  
 \usepackage[latin1]{inputenc}
 \usepackage{color}
 \usepackage{rotating}
 
\begin{document}
\preprint{LMU-ASC 40/07}
\title{On the Role of Zealotry in the Voter Model}
\author{M. Mobilia}
\affiliation{Arnold Sommerfeld Center for Theoretical Physics (ASC) and
Center for NanoScience (CeNS), 
Department of Physics, Ludwig-Maximilians-Universit\"at M\"unchen, 
Theresienstrasse 37, D-80333 M\"unchen, Germany}
\author{A. Petersen and S. Redner}
\affiliation{Center for Polymer Studies and Department of Physics, Boston University, Boston,
Massachusetts 02215 USA}
\date{\today}

\begin{abstract}
  We study the voter model with a finite density of zealots---voters that
  never change opinion.  For equal numbers of zealots of each species, the
  distribution of magnetization (opinions) is Gaussian in the mean-field
  limit as well as in one and two dimensions, with a width that is
  proportional to $1/\sqrt{Z}$, where $Z$ is the number of zealots,
  independent of the total number of voters.  Thus just a few zealots can
  prevent consensus or even the formation of a robust majority.

\end{abstract}

\maketitle

\section{Introduction}

The voter model \cite{L99} is one of the simplest examples of cooperative
behavior that has been used as a paradigm for the dynamics of opinions in
socially interacting populations.  In the voter model, each node of a graph
is occupied by a voter that has two opinion states, denoted as $+$ and $-$.
Opinions evolve by: (i) picking a random voter; (ii) the selected voter
adopts the state of a randomly-chosen neighbor; (iii) repeat these steps {\it
  ad infinitum} or until a finite system necessarily reaches consensus.
Naively, one can view each voter has having no self confidence and thus takes
on the state of one of its neighbors.  This evolution resembles that of the
Ising model with zero-temperature Glauber kinetics \cite{IG}, but with one
important difference: in the Ising model, each spin obeys the state of the
local majority; in the voter model, a voter chooses a state with a
probability that is proportional to the number of neighbors in that state.

There are three basic properties of the voter model that characterize its
evolution.  The first is the exit probability, namely, the probability that a
finite system eventually reaches consensus where all voters are in the +
state, $E_+(\rho_0)$, as a function of the initial density $\rho_0$ of +
voters.  Because the mean magnetization, defined as the difference in the
fraction of $+$ and $-$ voters (averaged over all realizations and
histories), is conserved on any degree-regular graph, and because the only
possible final states of a finite system are consensus, $E_+(\rho_0)=\rho_0$
\cite{L99}.

A second basic property is the mean time $T_N$ to reach consensus in a finite
system of $N$ voters.  For regular lattices in $d$ dimensions, it is known
that $T_N$ scales as $N^2$ in $d=1$, as $N\ln N$ in $d=2$, and as $N$ in
$d>2$ \cite{L99,K02}.  In contrast, $T_N$ generally scales sublinearly with
$N$ on heterogeneous graphs with broad degree distributions \cite{SR}.
Defining $\mu_k$ as the $k^{\rm th}$ moment of the degree distribution, then
$T_N\sim N\mu_1^2/\mu_2$, which grows slower than linearly in $N$ for a
sufficiently broad degree distribution.  Finally, the 2-point correlation
function $G_2(r)$, defined as the probability that 2 voters a distance $r$
apart are in the same state, asymptotically decays as $r^{2-d}$ on a regular
lattice when the spatial dimension $d>2$ \cite{K02,fpp}.  This decay is the
same as that of the electrostatic potential of a point charge, a
correspondence that has proven useful in analyzing the voter model.

\begin{figure}[ht]
  \includegraphics*[width=0.45\textwidth]{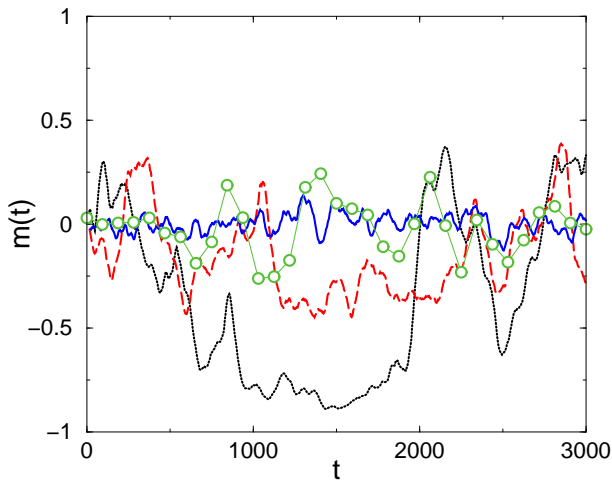}
  \caption{Time dependence of the magnetization for single realizations of
    1000 voters on the complete graph for $Z=2$ (black), 16 (red), and 128
    zealots (blue).  Data were smoothed over a 1\% range.  Also shown are
    U.S. presidential election results (circles) from 1876--2004
    (corresponding to $t=0$ and $t=3000$ respectively) where the
    magnetization is defined as the difference in the vote fraction of the
    top 2 candidates.}
  \label{mt}
\end{figure}

In this work, we investigate an extension of the voter model in which a small
fraction of the population are zealots---individuals that never change
opinion.  The effect of a single zealot \cite{1Z} or a small number of
zealots \cite{MG} on primarily static properties of the voter model has been
studied previously, and considerable insight has been gained by exploiting
the previously-mentioned electrostatic correspondence.  The role of zealots
has also been investigated in a majority rule opinion dynamics model
\cite{galam}, where again equal densities of zealots of each type prevent
consensus from being achieved.  One motivation for our work is the obvious
fact that consensus is not the asymptotic outcome of repeated elections in
democratic societies.  One such example is the set of US presidential
elections \cite{elections}, where the percentage of votes for the winner has
ranged from highs of 61.05\% (Johnson over Goldwater 1964) and 60.80\%
(Roosevelt over Landon 1932) to lows of 47.80\% (Harrison minority winner
over Cleveland 1888) and 47.92\% (Hayes over Tilden 1876).  In this
compilation, we exclude elections with substantial voting to candidates
outside the top two (Fig.~\ref{mt}).

This example, as well as election results from many democratic countries,
show in an obvious way that consensus will never be achieved in large voting
populations.  This fact motivates us to to investigate an opinion dynamics
model in which consensus is stymied by the presence of zealots.  Because of
the competing influences of the zealots and the tendency toward consensus by
the voter dynamics, the magnetization fluctuates with time in a manner that
can be made to qualitatively mimic, for example, the U.S.  presidential
election results (Fig.~\ref{mt}).  Upon averaging over a long time period,
these time-dependent fluctuations lead to a stationary magnetization
distribution whose properties are the main focus of this work.  

The basic question that we wish to address in the voter model with a
subpopulation of zealots is: what is the nature of the global opinion as a
function of the density of zealots?  One of our main results is that equal
but very small numbers of zealots of both types leads to a steady state with
a narrow Gaussian magnetization distribution centered at zero.  Here the
magnetization is simply the difference in the fraction of voters of each
species.  Thus a small fraction of zealots is surprisingly effective in
maintaining a steady state with only small fluctuations about this state.

It should also be mentioned that there are a variety of simple and
prototypical opinion dynamics models, in which lack of consensus is a basic
outcome, including the multiple-state Axelrod model \cite{A}, the bounded
compromise model of Weisbuch et al.\ \cite{W} and its variants \cite{VR}.
For these models, the consensus preventing feature typically is the absence
of interaction whenever two agents become sufficiently incompatible.  As a
function of basic model parameters, the fraction of incompatible agents can
grow, leading to cultural fragmentation and an attendant steady or static
opinion state.

In the next section, we define the model.  Then in Secs.~\ref{sec:cg} and
\ref{sec:1d}, we solve the model in the mean-field limit and on a
one-dimensional periodic ring.  We then investigate the behavior on the
square lattice by numerical simulations in Sec.~\ref{sec:2d} and find
behavior that is quantitatively close to that in the mean-field limit.
Finally, we conclude and point out some additional interesting features of
the role of zealotry on the voter model in Sec.~\ref{sec:conc}.

\section{The model}
\label{sec:model}

The population consists of $N$ voters, with a fixed number of zealots that
never change opinion, while the remaining voters are susceptible to opinion
change.  Each voter can be in one of two opinion states, $+1$ or $-1$ that we
term ``democrat'' and ``republican'', respectively.  Thus the system consists
of $Z_+$ democrat and $Z_-$ republican zealots, as well as $N_+$ democrat and
$N_-$ republican susceptibles.  Each type of voter evolves as follows:
 \begin{enumerate}
 \item {\it Susceptible democrats} can become republicans;
 \item {\it Susceptible republicans} can become democrats;
 \item {\it Zealot democrats} are always democrats;
 \item {\it Zealot republicans} are always republicans.
\end{enumerate}

Each agent, whether a zealot or a susceptible, has the same persuasion
strength that we set to $1$.  That is, after a susceptible voter selects a
neighbor, the voter is persuaded to adopt the state of this neighbor with
probability 1.  Because the total population comprises of agents in one of
four possible states, we have $N=N_+ + N_- + Z_+ + Z_-$.  Since the number of
zealots is fixed, the total number of susceptible individuals $S=
N-Z_+-Z_-=N_+ + N_-$ is also conserved.  The dynamics is a direct
generalization of voter model and consists of the following steps:
\begin{enumerate}
\item Pick a random voter, if this voter is a zealot nothing happens.
\item If the selected voter is a susceptible, then pick a random neighbor and
  adopt its state; note that if the selected voter and the neighbor are in
  the same state, nothing happens in the update.
\item Repeat steps 1 \& 2 {\it ad infinitum} or until consensus is reached.
\end{enumerate}
We will investigate this model on the two natural geometries of the complete
graph, a natural realization of the mean-field limit, and regular lattices.
For the complete graph, all other voters in the system are nearest-neighbor
to any voter.  Thus the complete graph has no spatial structure, a feature
that allows for a simple solution.  In contrast, when the voters live on the
sites of a regular lattice, a voter can be directly influenced only by its
the nearest neighbors.

\section{Dynamics on the complete graph}
\label{sec:cg}

On the complete graph, the state of the population may be characterized by
the probability $P(N_+, N_-,t)$ of finding $N_{\pm}$ susceptible voters at
time $t$.  Since $N_-=S-N_+$, we merely need to consider the master
equation for $P(N_+,t)$, which reads
\begin{eqnarray}
\label{ME}
\frac{\partial P(N_+,t)}{\partial t}&=&\sum_{\delta=\pm 1}
P(N_+ + \delta, t) W(N_+ + \delta \to N_+) 
\nonumber\\
&-&  \sum_{\delta=\pm 1} P(N_+, t) W(N_+ \to N_+ + \delta ).
\end{eqnarray}
The first term accounts for processes in which the number of susceptible
democrats after the event equals $N_+$, while the second term accounts for
the complementary loss processes where $N_+\to N_+\pm 1$.  Here $W$
represents the rate at which transitions occur and is given by
\begin{eqnarray}
\label{rates}
\delta t \;W(N_+ \to N_+ + 1) &=& \frac{N_-(N_+ +  Z_+)}{N(N-1)}
\nonumber\\
\delta t \;W(N_+ \to N_+ - 1) &=& \frac{N_+(N_- +  Z_-)}{N(N-1)}.
\end{eqnarray}
The first line is the probability of choosing first a republican susceptible
and then a democrat (susceptible or zealot), for which a susceptible
republican converts to a susceptible democrat in the voter model interaction.
We choose $\delta t = N^{-1}$, so that, on average, each agent is selected
once at each time step.

While it is usually not possible to solve an equation of the form
(\ref{ME}), analytical progress can be achieved by considering a continuum
$N\to\infty$ limit of the master equation and performing a Taylor expansion
\cite{Gard-VK}.  For this purpose, we introduce the rescaled variables
$n\equiv N_+/N$, $z_{\pm}=Z_{\pm}/N$, and also $s\equiv 1-z_+-z_-$ so that
$s-n\equiv N_-/N$.  In the continuum limit, the reaction rates now become
\begin{eqnarray}
\label{r}
W(n \to n + N^{-1}) &=& N\, (s-n)(n+ z_+)
\nonumber\\
W(n \to n - N^{-1}) &=& N\, n(s-n +  z_-).
\end{eqnarray}
Expanding (\ref{ME}) to the second order in the variable $n$, we find the
following Fokker-Planck equation \cite{EliPaul,VR,CST,RMF,McKane}:
\begin{eqnarray}
\label{FPE}
\frac{\partial P(n,t)}{\partial t}=-\frac{\partial}{\partial n} \left[\alpha(n) P(n,t)\right] + 
\frac{1}{2}\frac{\partial^2}{\partial n^2}\left[\beta(n) P(n,t)\right], 
\end{eqnarray}
where (see {\it e.g.}, Chap.~VII of Ref.~\cite{Gard-VK})
\begin{eqnarray*}
\label{alpha}
\alpha(n)&=& \sum_{\delta n= \pm 1/N} \delta n \;W(n \to n + \delta n)\\
&{}&~~~~= \left[z_+ s - n (1-s)\right];\\
\label{beta}
\beta(n) &=& \sum_{\delta n= \pm 1/N} (\delta n)^2 \;W(n \to n + \delta n) \\
&{}&~~~~= [(n+  z_+) (s-n)+n(s+ z_- - n)]/N.
\end{eqnarray*}

The first term on the right-hand-side of Eq.~(\ref{FPE}) leads to the
deterministic mean-field rate equation $\dot n(t)=
\alpha $, with solution
\begin{eqnarray}
\label{MF}
n(t)&=& \frac{ z_+ s}{ 1-s}+
 \left[n(0) - \frac{ z_+ s}{ 1-s}\right]\,
e^{-(1-s)t}.
\end{eqnarray}
Thus an initial density of susceptible democrats in an infinite system
exponentially relaxes to the steady-state value $n^*= z_+s/(1-s)$.
Correspondingly, the magnetization $m=(N_++Z_+-N_--Z_-)/N$ attains the
steady-state value $(z_+-z_-)/(z_++z_-)$.  When the number of agents is
finite, however, finite-size fluctuations arise from the diffusive second
term on the right-hand side of Eq.~(\ref{FPE}).  This term leads to a
steady-state probability distribution with a finite width that is centered at
$n^*$.  In what follows, we examine these fluctuations around the
mean-field steady state when $N$ and $Z_{\pm}$ are both finite.

\subsection{Stationary Magnetization Distribution}

According to the Fokker-Planck equation (\ref{FPE}), the stationary
distribution $P(n)$ obeys
\begin{eqnarray}
\label{FPE-stat}
\alpha(n) P(n) - \frac{1}{2}\frac{\partial}{\partial n}\left[\beta(n) P(n,t)\right]=0, 
\end{eqnarray}
whose formal solution is
\begin{eqnarray}
\label{Pstat}
P(n)= {\cal Z}\; \frac{{\rm exp}\left(2 \int_{0}^{n} dn'\, 
\frac{\alpha(n')}{\beta(n')} \right)}{\beta(n)}.
\end{eqnarray}
Since the density $n$ of agents in the state $+1$ ranges from $0$ to $s$, the
normalization constant ${\cal Z}$ is obtained by requiring $\int_{0}^{s} dn
P(n)= 1$.  This condition gives
\begin{equation*}
{\cal Z}=\left[\int_{0}^{s} \frac{{\rm
    exp}\left(2 \int_{0}^{n} dn'\, \frac{\alpha(n')}{\beta(n')}
  \right)}{\beta(n)}\, dn \right]^{-1}.
\end{equation*}

We are particularly interested in the distribution of the magnetization
$P(m)$ in the continuum limit, which directly follows from (\ref{Pstat})
through a simple change of variables.  We first consider the system with the
same number of zealots of each type, and then the asymmetric system with
unequal numbers of zealots of each type.

\subsection{Symmetric case: $Z_+=Z_-=Z$}

When the number of zealots of each species is equal, we write $Z_+=Z_-\equiv
Z$.  The rate equation (\ref{MF}) then gives an equal steady-state density of
democrats and republicans, $n^*=n_+=n_-=s/2$, corresponding to zero average
magnetization, $m^*=0$.  We now compute the stationary distribution of
magnetization by accounting for finite-size fluctuations.  When $Z_+=Z_-=Z$,
$P(n)$ obeys Eq.~(\ref{Pstat}) with
\begin{eqnarray*}
\label{a}
\alpha(n) &=& z (1-2z-2n), \\
\label{b} 
\beta(n) &=& [(2n+z)(1-2z)-2n^2]/N.
\end{eqnarray*}
Notice that $\alpha = \frac{Nz}{2}\frac{d\beta}{dn}$, a feature that allows
us to solve for the steady-state magnetization distribution easily.

\begin{figure}[ht]
  \includegraphics*[width=0.425\textwidth]{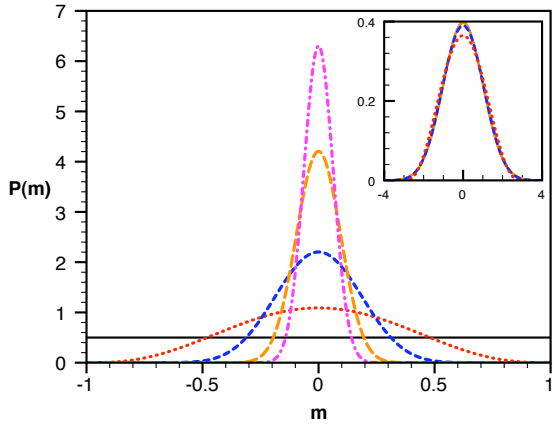}
  \caption{Steady-state magnetization distributions for 1000 voters on the
    complete graph for $Z=2,8,32,128$, and 512 zealots (progressively
    steepening curves).  The inset shows the scaled form of these
    distributions for $Z\geq 8$; the case $Z=8$ slightly deviates from the
    rest of the distributions that become visually coincident.}
  \label{zmf}
\end{figure}

To perform the integral in Eq.~\eqref{Pstat}, it is helpful to transform from
$n$ to the magnetization $m=(2n-s)/s$ which lies in $[-1,1]$.  We therefore
find ${\rm exp}\left(2\int_{0}^{n} dn' \frac{\alpha(n')}{\beta(n')}\right)=
\left(1+ \frac{2n(s-n)}{ z s}\right)^{ Nz}$.  According to Eq.~(\ref{Pstat}),
this leads to the following stationary distribution of susceptible democrats:
\begin{eqnarray}
\label{Pn}
P(n)&=& \frac{\left( z s + 2n(s-n)\right)^{ Nz -1}}
{\int_{0}^{s} dn\, \left( z s + 2n(s-n)\right)^{ Nz -1}}.
\end{eqnarray}
Using the fact that $2n(s-n)= s^2(1-m^2)/2$, we readily obtain the
stationary magnetization distribution:
\begin{eqnarray}
\label{Pm}
P(m)&=& \frac{\left( s^{-1} -m^2\right)^{ Z -1}}{\int_{-1}^{1} 
dm\, 
\left(s^{-1} -m^2\right)^{ Z -1}}.
\end{eqnarray}
In the limit of large $Z$, we may then approximate the distribution by the
Gaussian $P(m)\propto e^{-m^2/2\sigma^2}$, with $\sigma^2=/[2s(Z-1)]$.

When zealots are present in equal numbers, the magnetization distribution
quickly approaches a symmetric Gaussian, with a width that is inversely
proportional to the square-root of the {\em number} of zealots and not the
density.  Thus as the system size is increased, the density of zealots needed
to keep the magnetization within a fixed range goes to zero.  In the limiting
case where there is one zealot of each type, the magnetization is uniformly
distributed in $[-1,1]$ (Fig.~\ref{zmf}).  Finally notice that the
distribution quickly approaches the asymptotic scaling form when $Z\agt 8$
(inset to Fig.~\ref{zmf}).

\subsection{Asymmetric case: $Z_+\neq Z_-$}
\begin{figure}[ht]
\includegraphics*[width=0.425\textwidth]{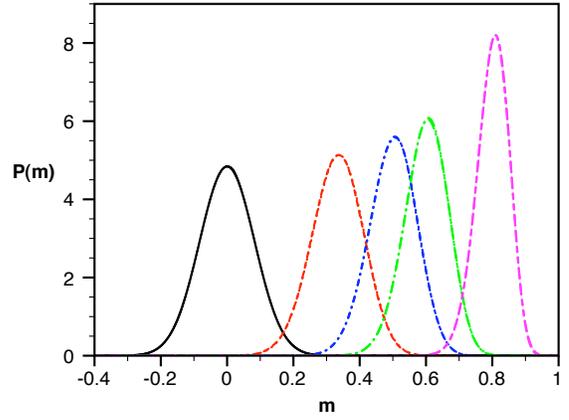}
\caption{Steady-state magnetization distributions on a complete graph of 1000
  sites with unequal numbers of zealots.  Shown left to right are the cases
  of $(Z_+,Z_-)=(90,90)$, $(120,60)$, $(135,45)$, $(144,36)$, $(162,18)$.
  The results of voter model simulations and the solution to the master
  equations are coincident.  The mean magnetization of the system equals the
  magnetization of the zealots: $m= \frac{z_+-z_-}{z_++z_-} $. }
\label{zmf-asymm}
\end{figure}
When the density of zealots of each type are unequal, we now have
\begin{eqnarray}
\label{a2}
\alpha(n) &=& (z_+ +n)s -n, \\
\label{b2} 
\beta(n) &=& [(2n+z_+) ( s-n) + nz_- ]/N
\end{eqnarray}
in Eq.~(\ref{FPE-stat}).  To compute $P(n)$ (and equivalently $P(m)$), it is
now convenient to introduce the quantities $\delta\equiv z_+ - z_-$ and
$r\equiv \sqrt{\delta^2 + 4s}$.  Noticing that one can write $\alpha/\beta=
[N(s-1)(d\beta/dn)+\delta(1+s)/4]/\beta$,  one can easily
compute the integral in Eq.~\eqref{Pstat} and thereby obtain $P(n)$.
Transforming from the density to the magnetization by $n=(m+1)s/2$, we obtain
the following expression for the stationary magnetization distribution
(Fig.~\ref{zmf-asymm}):
\begin{eqnarray}
\label{Pm_2}
{\cal Z}P(m)&=& \left[1 -m (\delta + ms)\right]^{(Z_+ + Z_- -2)/2} 
\nonumber\\
&\times& \left[ 1+\frac{r}{ms -\frac{r-\delta}{2}}\right]
^{(\delta/2r)\,(2N-Z_+-Z_-)}.\nonumber\\
\end{eqnarray}

As in the symmetric case, ${\cal Z}$ is a normalization constant obtained by
requiring that $\int_{-1}^{1} dm\, P(m)=1$.  Notice that $P(m)$ is comprised
of two terms.  The first term gives a Gaussian contribution (in the limit of
large $N$) and is the analog of Eq.~(\ref{Pm}).  The second term is a
nontrivial contribution due to the asymmetry that is responsible for the
skewness of $P(m)$ which remains peaked around $m^*=\frac{z_+-z_-}{z_+ +
  z_-}$.  Close to this peak value, there is little asymmetry ({\it i.e.},
$\delta \ll 1)$.  Additionally, for a large number of zealots we may
approximate the distribution (\ref{Pm_2}) by the Gaussian $P(m)\approx
e^{-(m-m^*)^2/2\sigma^2}\,[1+{\cal O}((m-m^{*})\delta))]$, with
$\sigma^2=[c(Z_+ + Z_- -2)]^{-1}$.
\\
\section{ONE DIMENSION}
\label{sec:1d}

We now turn to the one-dimensional system, where the behavior of the
classical voter model is quite different from that in the mean-field limit.
When zealots are present, however, we generically obtain a Gaussian
magnetization distribution, as in the mean-field case.  We now derive the
magnetization distribution---first for two zealots---and then for an
arbitrary number of zealots.

\subsection{Two Zealots}

\begin{figure}[ht]
\includegraphics*[width=0.2250\textwidth]{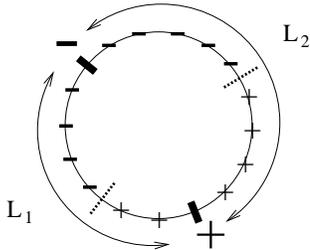}
\caption{A ring divided into two independent segments by oppositely-oriented
  zealots (thick lines).  Also shown is the state of each voter and the
  domain wall in each segment at long times (dotted lines). }
\label{ring}
\end{figure}

Suppose that two zealots of opposite opinion are randomly placed on a
periodic ring of length $L$.  The ring is thus split into two independent
segments of lengths $L_1$ and $L_2$, with $L=L_1+L_2+2$ (Fig.~\ref{ring}).
We take the ring to be large so that we may write $L\approx L_1+L_2$.  As
shown in Fig.~\ref{ring}, the voters in each segment coarsen and eventually
there remains one domain of $+$ voters that is separated from one domain of
$-$ voters by a single domain wall.  Each domain wall performs a free random
walk and the walk is reflected upon reaching the end of its segment.  A basic
fact from the theory of random walks \cite{Weiss} is that each domain wall is
equiprobably located within the interval in the long-time limit.  We now
exploit this property to determine the magnetization distribution.

\begin{figure}[ht]
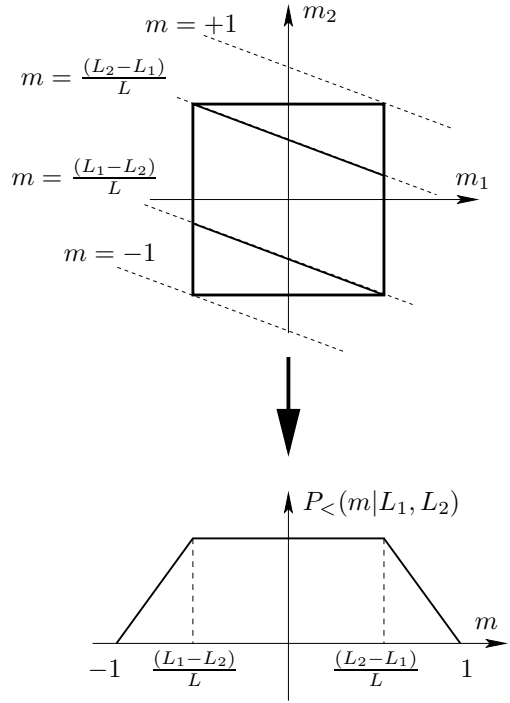

\input pm.pstex_t
\caption{(Top) Rays of fixed magnetization (dashed) for the case $L_1<L_2$.
  The probability for a given value of $m$ is proportional to the length of
  the ray corresponding to this $m$ value within the unit square (solid).
  (Bottom) The resulting magnetization distribution $P_<(m|L_1,L_2)$ for a
  given $L_1$ and $L_1<L_2$.}
\label{pm}
\end{figure}

For interval lengths $L_1$ and $L_2$ and respective magnetizations $m_1$ and
$m_2$, the magnetization $m$ of the entire ring is given by
$mL=m_1L_1+m_2L_2$.  Thus a given value of $m$ is achieved if $m_1$ and $m_2$
are related by (Fig.~\ref{pm})
\begin{equation}
\label{ray}
m_2=\frac{mL}{L_2}-\frac{m_1L_1}{L_2}.
\end{equation}
Then the probability $P(m|L_1,L_2)$ for a system of two segments with lengths
$L_1$ and $L_2$ to have magnetization equal to $m$ is proportional to the
length of the ray defined by Eq.~\eqref{ray} that lies within the unit square
in the $m_1$-$m_2$ plane.  As illustrated in Fig.~\ref{pm}, the distribution
$P_<(m|L_1,L_2)$, where the subscript $<$ now signifies the range $L_1<L_2$,
increases linearly with $m$ for $-1<m<(L_1-L_2)/L$, is constant for
$(L_1-L_2)/L<m<(L_2-L_1)/L$, and then decreases linearly with $m$ for
$(L_2-L_1)/L<m<1$.

Using this $m$ dependence of $P_<(m|L_1,L_2)$ and also imposing normalization,
we thus find the magnetization distribution for fixed $L_1, L_2$ with
$L_1<L_2$ to be:
\begin{equation}
\label{PmLi}
P_<(m|L_1,L_2)=
\begin{cases}
{\displaystyle \frac{L^2(1+m)}{4L_1L_2}} &~~~ -1<m<\frac{L_1-L_2}{L}\\ \\
{\displaystyle \frac{L}{2L_2}} & ~~~~~~~~~~~\,|m|<\frac{L_2-L_1}{L} \\ \\
{\displaystyle \frac{L^2(1-m)}{4L_1L_2}} & \,\frac{L_2-L_1}{L}<m<1.
\end{cases}
\end{equation}
The complementary distribution $P_>(m|L_1,L_2)$ for $L_1>L_2$ is obtained
from Eq.~\eqref{PmLi} by interchanging the roles of $L_1$ and $L_2$.

\begin{figure}[ht]
\includegraphics*[width=0.425\textwidth]{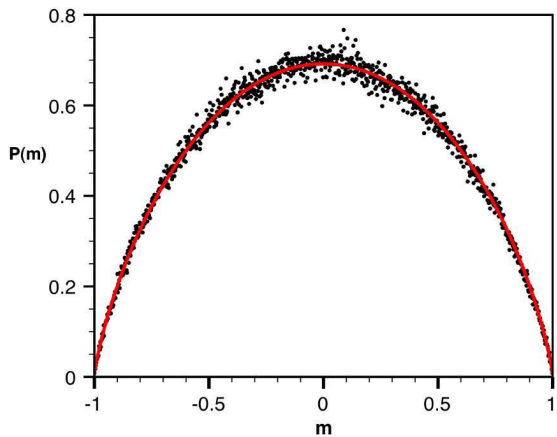}
\caption{Comparison the analytic magnetization distribution for two zealots
  on the ring (Eq.~\eqref{Pmav}) and simulation results (points). }
\label{ring-compare}
\end{figure}

Now we integrate over all values of $L_1$ to find the configuration-averaged
magnetization distribution $P(m)$.  The details of this calculation are a bit
tedious and are given in Appendix~\ref{app:pm}.  The final result is
\begin{eqnarray}
\label{Pmav}
P(m) &\!=\!& \frac{1}{L}\left[ \int_0^{\frac{L}{2}}\!\! P_<(m|L_1,L_2)\, dL_1
+  \int_{\frac{L}{2}}^L \!\!P_>(m|L_1,L_2)\, dL_1\right]\nonumber \\
&\!=\!& \left(\frac{1-|m|}{2}\right)\ln\left(\frac{1+|m|}{1-|m|}\right) - 
\ln\left(\frac{1+|m|}{2}\right).
\end{eqnarray}
As shown in Fig.~\ref{ring-compare}, the agreement between Eq.~\eqref{Pmav}
and simulations is excellent.

\subsection{Many Zealots}

We now study the magnetization distribution when many zealots are randomly
distributed on the ring, with the restriction of equal numbers of each type
of zealot ($Z_+=Z_-=Z$).  Two distinct possibilities can arise:

\begin{enumerate}
\item A segment of consecutive susceptible voters is surrounded by two
  zealots of the same sign.  With voter model dynamics, these segments
  eventually align with the state of the confining zealots so that the
  segment freezes.
\item A segment of consecutive susceptible voters is surrounded by two
  zealots of opposite opinion.  Eventually a single domain wall remains that
  diffuses freely within the segment.
\end{enumerate}
We first consider the simpler case where equal numbers of + and $-$ zealots
are randomly but alternately placed around the ring so that no frozen
segments arise.  The segment lengths $\lbrace L_{i}\rbrace$ with
$i=1,2,\ldots,Z$, obey the constraint $\sum_i L_i=L$ (ignoring the space
occupied by the zealots themselves).

To find the magnetization distribution, we map the state of the voters onto
an equivalent random walk as follows.  In a segment of length $L_i$, the
difference in the number of $+$ and $-$ voters at long times is uniformly
distributed in $[-L_i,L_i]$.  We define this difference as the unnormalized
magnetization $M_i$.  We now make the following approximations that apply
when $L,Z\to\infty$ such that each $L_i$ is also large.  In this limit, we
may assume that each $L_i$ is independent and identically distributed.  As a
result, the sum of the unnormalized magnetizations over all intervals is
equivalent to the displacement of a random walk of $Z$ steps with each step
uniformly distributed in $[-L_i,L_i]$.

To solve this random walk problem, we use the basic fact that the Fourier
transform for the probability distribution of the entire walk
$\mathcal{P}(k)$ is simply the product of the Fourier transforms of the
single-step distributions \cite{fpp,Weiss}.  Since the Fourier transform of a
uniform single-step distribution over the range $[-L_i,L_i]$ is $\frac{sin
  kL_i}{kL_i}$, we then have
\begin{equation}
\label{Pk}
\mathcal{P}(k) =\prod_{i=1}^Z \frac{\sin kL_i}{kL_i}.
\end{equation}

Since we are interested in the asymptotic limit where the unnormalized
magnetization becomes large, we study the limit of  $\mathcal{P}(k)$ for
small $k$.  Thus we expand  each factor in $\mathcal{P}(k)$ in a Taylor
series to first order, and then re-exponentiate to yield
\begin{eqnarray*}
\mathcal{P}(k)&\approx& \prod_{i=1}^Z (1 -k^2L_i^2/6)\\
&\sim& 1 -\sum_i^Z k^2L_i^2/6 \sim e^{-k^2\sum_i L_i^2/6}.
\end{eqnarray*}
We now invert this Fourier transform to give the distribution of the
unnormalized magnetization
\begin{eqnarray}
\label{PM}
P(M)&=&\frac{1}{2\pi}\int  e^{-k^2\sum_i L_i^2/6}\, e^{-ikM}\, dk\nonumber \\
&=& \frac{1}{\sqrt{2\pi \sigma_M^2}}\,\, e^{-M^2/2\sigma_M^2},
\end{eqnarray}
with $\sigma_M^2 =\sum_iL_i^2/3$.

What we want, however, is the magnetization distribution; this is related to
$P(M)$ by $P(m)\,dm=P(M)\,dM$.  We thus find
\begin{equation}
\label{Pm-final}
P(m)= \frac{1}{\sqrt{2\pi \sigma_m^2}}\,\, e^{-m^2/2\sigma_m^2},
\end{equation}
where $\sigma_m^2 =\sum_iL_i^2/3L^2$.  If the number of intervals is large,
then each $L_i$ is approximately $L/Z$, from which we obtain
$\sigma_m^2\approx 1/3Z$.  (The result $\sigma_m^2=1/3Z$ is exact if all
interval lengths are equal.)~ As in the mean-field limit, the width of the
magnetization distribution is controlled by the {\em number} of zealots and
not their concentration, so that a small number of zealots is effective in
maintaining the magnetization close to zero.

A similar approach applies in the case where the spatial ordering of the
zealots is uncorrelated.  In this case, approximately one-half of all
segments will be terminated by oppositely-oriented zealots and one-half by
zealots of the same species.  For the latter type of segments, the
unnormalized magnetization will equal $\pm L_i$ equiprobably.  Under the
assumption that exactly one-half of the segments are frozen and one-half
contain a single freely-diffusing domain wall, the analog of Eq.~\eqref{Pk}
is
\begin{equation}
\label{Pk-gen}
\mathcal{P}(k) =\prod_{i=1}^{Z/2} \frac{\sin kL_i}{kL_i}\,\,\,\, \prod_{i=1}^{Z/2} \cos kL_i.
\end{equation}
The second product accounts for frozen segments in which the unnormalized
magnetization equals $\pm L_i$ equiprobably.  For these segments the Fourier
transform of the single-step probability for a random walk whose steps length
are $\pm L_i$ equals $\cos kL_i$.  Following the same steps that led to
Eq.~\eqref{Pm-final}, we again obtain a Gaussian magnetization distribution,
but with $\sigma_m^2$ given by $\sigma_m^2 =\sum_i2L_i^2/3L^2\to 2/3Z$.

\section{Two Dimensions}
\label{sec:2d}

\begin{figure}[ht]
\includegraphics*[width=0.425\textwidth]{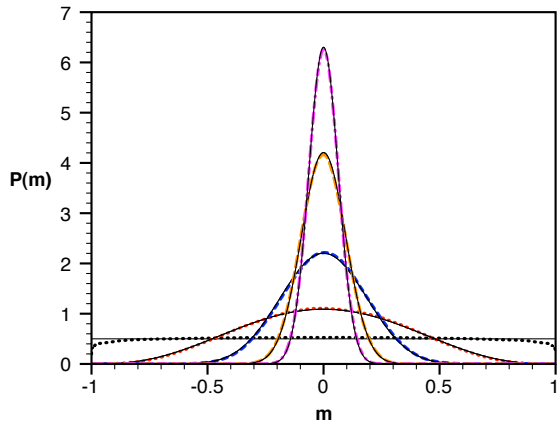}
\caption{Comparison of simulations for the magnetization distribution in two
  dimensions (dashed) with the mean-field results (solid curves).  The
  simulations are for 1000 voters with $2,8,32,128$ and 512 total zealots,
  with equal numbers of each type.}
\label{tsymm}
\end{figure}

In the classical voter model, the two-dimensional system is at the critical
dimension so that its behavior deviates from that of the mean-field system by
logarithmic corrections.  In the presence of zealots, however, the behaviors
in two dimensions and in mean field are quite close, as illustrated in
Fig.~\ref{tsymm}.  

Our results for two dimensions are based on numerical simulations.  In our
simulations, we pick a random voter and apply the update rules of
Sec.~\ref{sec:model}.  The unit of time is defined so that a time increment
$dt=1$ corresponds to $N$ update events, so that each voter is updated once
on average.  The system is initialized with each voter equally likely to be
in the $+$ or the $-$ states.  From the $N_+$ voters in the $+$ state, $Z_+$
of them are designated as zealots, and similarly for voters in the $-$ state.
After the system reaches the steady state, we measure steady-state properties
at time intervals $\Delta T$.  The delay time $T$ to reach the steady state
depends on the lattice dimension and the zealot density, while $\Delta T$ is
the correlation time for the system in the steady state.  By making
measurements every $\Delta T$ steps, we obtain data for effectively
uncorrelated systems.  Typically, for a given initial condition, we made 100
measurements and then averaged over many configurations of zealots.

The resulting data for the magnetization distribution is typically noisy, and
we employ a Gaussian averaging of nearby points to smooth the data.  If
$m_i$ denotes the $i^{\rm th}$ magnetization value, then the smoothed
magnetization distribution at $m_i$ is defined as
\begin{equation*}
\overline{P(m_i)} = \frac{1}{\sqrt{\pi d^2}}\sum_{k=-d}^{d}P(m_{i+k})\, e^{-(k/d)^2},
\end{equation*}
where the sum includes the initial point, as well as the $d$ points to the
left and to the right of the initial point, with $d$ typically in the range
20--40.  Such a smoothed distribution is the quantity that is actually
plotted in Figs.~\ref{zmf}, \ref{zmf-asymm}, \ref{tsymm}, and in the spatial
averaged distribution in Fig.~\ref{2ddipole}.

\section{Discussion}
\label{sec:conc}

We have shown that a small number of zealots in a population of voters is
quite effective in maintaining a steady state in which consensus is never
achieved.  When there are equal numbers of zealots of each type, the
steady-state fraction of democrats and republicans equals 1/2; equivalently,
the magnetization equals zero.  For unequal densities of the two types of
zealots, the steady-state magnetization is simply $m^*=(Z_+-Z_-)/(Z_++Z_-)$,
where $Z_+$ and $Z_-$ are the number of zealots of each type.  The
magnetization distribution is generically Gaussian, $P(m)\propto
e^{-(m-m^*)^2/2\sigma^2}$, with $\sigma\propto 1/\sqrt{Z}$, and $Z=Z_++Z_-$
is the total number of zealots.  A Gaussian magnetization distribution arises
universally in one dimension, on the square lattice (two dimensions), and on
the complete graph (mean-field limit).  One basic consequence of this
distribution is that as the total number of voters $N$ increases, the
fraction of zealots needed to keep the magnetization less than a specified
level vanishes as $1/\sqrt{N}$. 

There are several additional aspects of the influence that zealots have on
the voter model that are worth pointing out.  Although the time to reach
consensus is infinite because this state can never be achieved, one can ask
for the time until a specified plurality is first achieved.  Equivalently, we
can ask for the probability that the magnetization first reaches a value $m$,
when the system is initialized with $m=m_0$.  From the above generic Gaussian
form of the magnetization distribution, we expect that the mean time for a
symmetric system to first reach a magnetization $m$ will thus scale as
$e^{am^2Z}$, where $a$ is a constant of order one.  Thus one must wait an
extremely long time before the system achieves even a modest deviation away
from the zero-magnetization state when the number of zealots becomes
appreciable.  Perhaps this trivial fact is the underlying reason why so many
democratic countries are characterized by small majorities in governance.

\begin{figure}[ht]
\includegraphics*[width=0.425\textwidth]{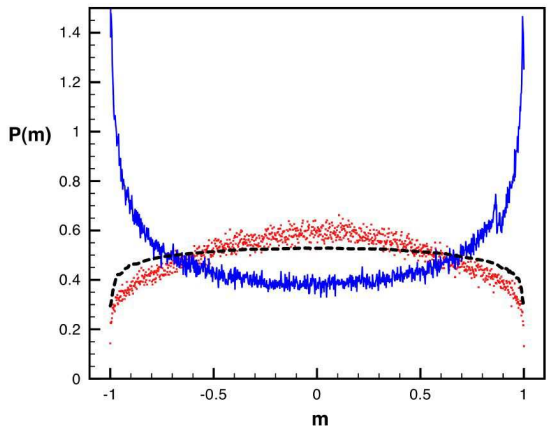}
\caption{Comparison of simulations for the magnetization distribution in two
  dimensions when the two zealots are adjacent (curve with peaks near $\pm
  1$), maximally separated (dots), and averaged over many different zealot
  configurations (dashed). }
\label{2ddipole}
\end{figure}

Another interesting feature is the role of the zealots' spatial positions on
the steady state.  For example, if there are only two zealots that are
adjacent, one might expect that the effect of this ``dipole'' would be weaker
than that of two separated monopoles.  This is precisely the effect that is
is observed in Fig.~\ref{2ddipole}.  When the two zealots are adjacent, their
effects are substantially screened and the magnetization distribution is
peaked near $m=\pm 1$.  That is, the voters show a preference for consensus
in spite of the zealots.  On the other hand, when the zealots are maximally
separated, the magnetization distribution is close to the distribution that
arises when averaging over possible positions of the two zealots.

Zealots are also quite effective in reducing the total number of opinion
changes in the system.  If the population is close to zero magnetization,
each voter typically has equal numbers of neighbors of each type.  If the
voters are not strongly correlated, each voter would change its state at a
rate that is approximately equal to 1/2.  However, simulations on the square
lattice show that the flip rate of each susceptible voter is considerably
smaller.  For example, for 1000 voters with 10 zealots (5 of each type), the
rate of opinion changes of the susceptibles is around 1/5 and this rate
decreases as the density of zealots decreases.

Finally, a slight embellishment of our model could apply to real voting
patterns in a democracy with strong regional differences.  Here it is natural
to partition a population into enclaves, with an imbalance of one type of
zealot over the other in each enclave.  Such a spatial distribution would
correspond to red (republican) and blue (democrat) states in the parlance of
US electoral politics.  It would be interesting to study if such an extension
can actually account for election results.

{\it Added Note:} As this manuscript was being written, we became aware of a
very recent eprint by Chinellato et al.\ \cite{necsi}; they study essentially
the same model as in this work, but with a somewhat different focus than
ours.

\acknowledgments We gratefully acknowledge the support of the Alexander von
Humboldt Foundation through the fellowship IV-SCZ/1119205 (MM) and NSF grant
DMR0535503 (AP and SR).

\begin{widetext}

\appendix*

\section{Magnetization distribution for two zealots}
\label{app:pm}

We want to compute the integral
\begin{equation}
P(m) = \frac{1}{L}\left[ \int_0^{\frac{L}{2}} P_<(m|L_1,L_2)\, dL_1
+  \int_{\frac{L}{2}}^L P_>(m|L_1,L_2)\, dL_1\right].
\end{equation}
Since $P_<(m|L_1,L_2)$ and $P_>(m|L_1,L_2)$ have different forms in different
parts of the interval $[-1,1]$, each of the above integrals needs to be split
into two parts.  For $P_<(m|L_1,L_2)$ and assuming that $m>0$, the linear
ramp part of the probability distribution needs to be used for
$(L_2-L_1)/L<m$, which translates for $L_1>L(1-m)/2$.  Similarly, for
$P_>(m|L_1,L_2)$ and again for $m>0$, the linear ramp must be used when
$(L_1-L_2)/L<m$, or $L_1<L(1+m)/2$.  Thus the above integral becomes
\begin{equation}
P(m) = \frac{1}{L}\left[ \int_0^{\frac{L}{2}(1-m)} \frac{L\, dL_1}{2(L-L_1)}
+  \int_{\frac{L}{2}(1-m)}^{\frac{L}{2}} \frac{(1-m)L^2\, dL_1}{4L_1(L-L_1)}\,
+  \int_{\frac{L}{2}}^{\frac{L}{2}(1+m)} \frac{(1-m)L^2\, dL_1}{4L_1(L-L_1)}\,
+\int_{\frac{L}{2}(1+m)}^L \frac{L\, dL_1}{2L_1}\right].
\end{equation}
Each of these integrals is then elementary.  We also obtain the result for
$m<0$ by reflecting the result of the above integral about $m=0$ to give
Eq.~\eqref{Pmav}.

\end{widetext}

\end{document}

%% file: pm.pstex_t
\begin{picture}(0,0)%
\includegraphics{pm.pstex}%
\end{picture}%
\setlength{\unitlength}{1579sp}%
\begingroup\makeatletter\ifx\SetFigFont\undefined%
\gdef\SetFigFont#1#2#3#4#5{%
  \reset@font\fontsize{#1}{#2pt}%
  \fontfamily{#3}\fontseries{#4}\fontshape{#5}%
  \selectfont}%
\fi\endgroup%
\begin{picture}(8036,11035)(451,-10283)
\put(1276,-3361){\makebox(0,0)[lb]{\smash{{\SetFigFont{10}{12.0}{\familydefault}{\mddefault}{\updefault}{\color[rgb]{0,0,0}$m=-1$}%
}}}}
\put(2551,239){\makebox(0,0)[lb]{\smash{{\SetFigFont{10}{12.0}{\familydefault}{\mddefault}{\updefault}{\color[rgb]{0,0,0}$m=+1$}%
}}}}
\put(451,-2161){\makebox(0,0)[lb]{\smash{{\SetFigFont{10}{12.0}{\familydefault}{\mddefault}{\updefault}{\color[rgb]{0,0,0}$m=\frac{(L_1-L_2)}{L}$}%
}}}}
\put(601,-586){\makebox(0,0)[lb]{\smash{{\SetFigFont{10}{12.0}{\familydefault}{\mddefault}{\updefault}{\color[rgb]{0,0,0}$m=\frac{(L_2-L_1)}{L}$}%
}}}}
\put(2626,-9886){\makebox(0,0)[lb]{\smash{{\SetFigFont{10}{12.0}{\familydefault}{\mddefault}{\updefault}{\color[rgb]{0,0,0}$\frac{(L_1-L_2)}{L}$}%
}}}}
\put(1651,-9886){\makebox(0,0)[lb]{\smash{{\SetFigFont{10}{12.0}{\familydefault}{\mddefault}{\updefault}{\color[rgb]{0,0,0}$-1$}%
}}}}
\put(7501,-9886){\makebox(0,0)[lb]{\smash{{\SetFigFont{10}{12.0}{\familydefault}{\mddefault}{\updefault}{\color[rgb]{0,0,0}$1$}%
}}}}
\put(5476,-9886){\makebox(0,0)[lb]{\smash{{\SetFigFont{10}{12.0}{\familydefault}{\mddefault}{\updefault}{\color[rgb]{0,0,0}$\frac{(L_2-L_1)}{L}$}%
}}}}
\put(7726,-9136){\makebox(0,0)[lb]{\smash{{\SetFigFont{10}{12.0}{\familydefault}{\mddefault}{\updefault}{\color[rgb]{0,0,0}$m$}%
}}}}
\put(7426,-2161){\makebox(0,0)[lb]{\smash{{\SetFigFont{10}{12.0}{\familydefault}{\mddefault}{\updefault}{\color[rgb]{0,0,0}$m_1$}%
}}}}
\put(5026,464){\makebox(0,0)[lb]{\smash{{\SetFigFont{10}{12.0}{\familydefault}{\mddefault}{\updefault}{\color[rgb]{0,0,0}$m_2$}%
}}}}
\put(5026,-7261){\makebox(0,0)[lb]{\smash{{\SetFigFont{10}{12.0}{\familydefault}{\mddefault}{\updefault}{\color[rgb]{0,0,0}$P_<(m|L_1,L_2)$}%
}}}}
\end{picture}%